\documentclass[twocolumn,prl,aps,showpacs,footinbib,amsmath,amssymb,superscriptaddress,longbibliography]{revtex4-1}
\usepackage{graphicx}
\usepackage{dcolumn}
\usepackage{xcolor}
\usepackage{bm}
\usepackage{hyperref}
\usepackage{amsmath,amssymb,amsfonts,bbold}
\usepackage[normalem]{ulem}

\newcommand{\beq}{\begin{equation}}
\newcommand{\eeq}{\end{equation}}
\newcommand{\beqa}{\begin{eqnarray}}
\newcommand{\eeqa}{\end{eqnarray}}

\graphicspath{{fig/}}
\usepackage[normalem]{ulem}

\begin{document}
\title{Anisotropic vortices on superconducting Nb(110)}
\author{Artem~Odobesko}
	\email[corresponding author: \\]{artem.odobesko@physik.uni-wuerzburg.de}
	\affiliation{Physikalisches Institut, Experimentelle Physik II,
		Universit\"{a}t W\"{u}rzburg, Am Hubland, 97074 W\"{u}rzburg, Germany}
\author{Felix Friedrich}
	\affiliation{Physikalisches Institut, Experimentelle Physik II,
		Universit\"{a}t W\"{u}rzburg, Am Hubland, 97074 W\"{u}rzburg, Germany}
\author{Song-Bo Zhang}
	\affiliation{Theoretische Physik IV, Institut f{\"u}r Theoretische Physik und Astrophysik, 
		Universit\"{a}t W\"{u}rzburg, Am Hubland, 97074 W\"{u}rzburg, Germany}
\author{Soumyajyoti Haldar}
	\affiliation{Institut f\"{u}r Theoretische Physik und Astrophysik, 
		Christian-Albrechts-Universit\"{a}t zu Kiel, Leibnizstr. 15, 24098 Kiel, Germany}
\author{Stefan Heinze}
	\affiliation{Institut f\"{u}r Theoretische Physik und Astrophysik, 
		Christian-Albrechts-Universit\"{a}t zu Kiel, Leibnizstr. 15, 24098 Kiel, Germany}
\author{Bj\"{o}rn Trauzettel}
	\affiliation{Theoretische Physik IV, Institut f{\"u}r Theoretische Physik und Astrophysik, 
		Universit\"{a}t W\"{u}rzburg, Am Hubland, 97074 W\"{u}rzburg, Germany}
    \affiliation{W\"urzburg-Dresden Cluster of Excellence ct.qmat, Germany}
\author{Matthias Bode}
	\address{Physikalisches Institut, Experimentelle Physik II,
	Universit\"{a}t W\"{u}rzburg, Am Hubland, 97074 W\"{u}rzburg, Germany}
	\affiliation{W\"urzburg-Dresden Cluster of Excellence ct.qmat, Germany}
	\address{Wilhelm Conrad R{\"o}ntgen-Center for Complex Material Systems (RCCM),
	Universit\"{a}t W\"{u}rzburg, Am Hubland, 97074 W\"{u}rzburg, Germany}

\begin{abstract}
We investigate the electronic properties of type-II superconducting Nb(110) in an external magnetic field.
Scanning tunneling spectroscopy reveals a complex vortex shape   
which develops from circular via coffee bean-shaped to elliptical 
when decreasing the energy from the edge of the superconducting gap to the Fermi level.
This anisotropy is traced back to the local density of states of Caroli--de Gennes--Matricon states 
which exhibits a direction-dependent splitting.
Oxidizing the Nb(110) surface triggers the transition from the clean to the dirty limit, 
quenches the vortex bound states, and leads to an isotropic appearance of the vortices.
Density functional theory shows that the Nb(110) Fermi surface 
is stadium-shaped near the $\overline{\Gamma}$ point.
Calculations within the Bogoliubov-de-Gennes theory using these Fermi contours 
consistently reproduce the experimental results.
\end{abstract}
\maketitle

{\em Introduction---}
In the recent past, we have witnessed a growing interest in the physical properties
of vortices in unconventional and topological superconductors,
as they allow for the emergence of  Majorana zero modes (MZMs) \cite{Fischer2007, Ivanov2001, Lutchyn2010}
which may potentially be used for applications in quantum computation \cite{Nayak2008}.
Although some success has been achieved on heavy electron iron-based superconductors \cite{Chen2020, Kong2019},
the unambiguous spectroscopic identification of MZMs remains demanding,
as their spectroscopic distinction from trivial quasiparticle excitations 
at experimentally accessible temperatures still constitutes significant challenges~\cite{Yuan2019, Yin2015, Sun2016}.

In vortices, such trivial excitations occur when the electron mean free path $\ell$ is much larger
than the vortex diameter, i.e., the superconducting coherence length $\xi$.
As predicted by Caroli, de Gennes, and Matricon in the 1960s \cite{Caroli1964},
Andreev reflection inside the vortex gives rise to a discrete set of bound states.
Their separation in energy is given by $\Delta^2/E_{\rm{F}}$, 
where $2 \Delta$ is the width of the superconducting gap and $E_{\rm{F}}$ is the Fermi energy,
often resulting in values in the $\mu$eV range only~\cite{Hayashi1998}.
At such low splittings, the CdGM states above and below the Fermi level can hardly be resolved
even if measurements are performed at milli-Kelvin temperatures,
as they thermally merge into one broad peak which appears to be energetically positioned
at zero bias \cite{Xu2015,Liu2018,Machida2019}.

Furthermore, the spatial distribution of specific CdGM states inside the vortex 
scales with their angular momentum $\mu$, resulting in a wave function which peaks 
at a distance $r_{\mu} \approx |\mu|/k_F$ away from the vortex core \cite{Shore1989, Hayashi1998}.
However, the specific spatial and energy distribution of CdGM bound states in a given material 
depends on various parameters, such as the pairing anisotropy, spin-orbit coupling, Fermi surface anisotropy, 
or vortex--vortex interactions \cite{Hayashi1996, Gygi1991, Melnikov2009}.
In fact, the resulting patterns of vortex bound states can be quite complex \cite{Hess1990, Hess1991}.
Furthermore, in the dirty limit ($\ell \lesssim \xi$), scattering processes lead to energetic broadening 
and eventually the complete quenching of the CdGM states \cite{Renner1991}. 
Thus, detailed investigations of the electronic structure inside superconducting vortices
can help to better understand and tell apart trivial from topological states.

In this paper, we present results of a scanning tunneling spectroscopy (STS) study of vortex bound states
on clean and oxygen-reconstructed Nb(110) in an external out-of-plane magnetic field.
Whereas no bound states are found on the oxygen-reconstructed surface,
a well pronounced zero-bias peak appears in the vortex core of clean Nb(110).
Differential conductance $\mathrm{d}I/\mathrm{d}U$ maps measured at constant tip--sample distance reveal
that the spatial distribution of the local density of states (LDOS) crucially depends on the bias voltage.
While round-shaped vortices are visualized when the bias is set close to the edge of the superconducting gap,
a two-fold anisotropy is observed at zero bias with an intermediate coffee bean-shaped form 
(the shape of a roasted {\em coffea robusta} bean, to be precise).
The elliptical shape of the LDOS maps is qualitatively explained by the anisotropy of the Fermi surface
which exhibits a stadium-like shape around the $\overline{\Gamma}$ point of the surface Brillouin zone (SBZ).
Using this anisotropic Fermi surface as an input, the experimental data are modeled
by solving self-consistently the Bogoliubov-de Gennes (BdG) equation.


\begin{figure*}[t]
	\includegraphics[width=1\textwidth]{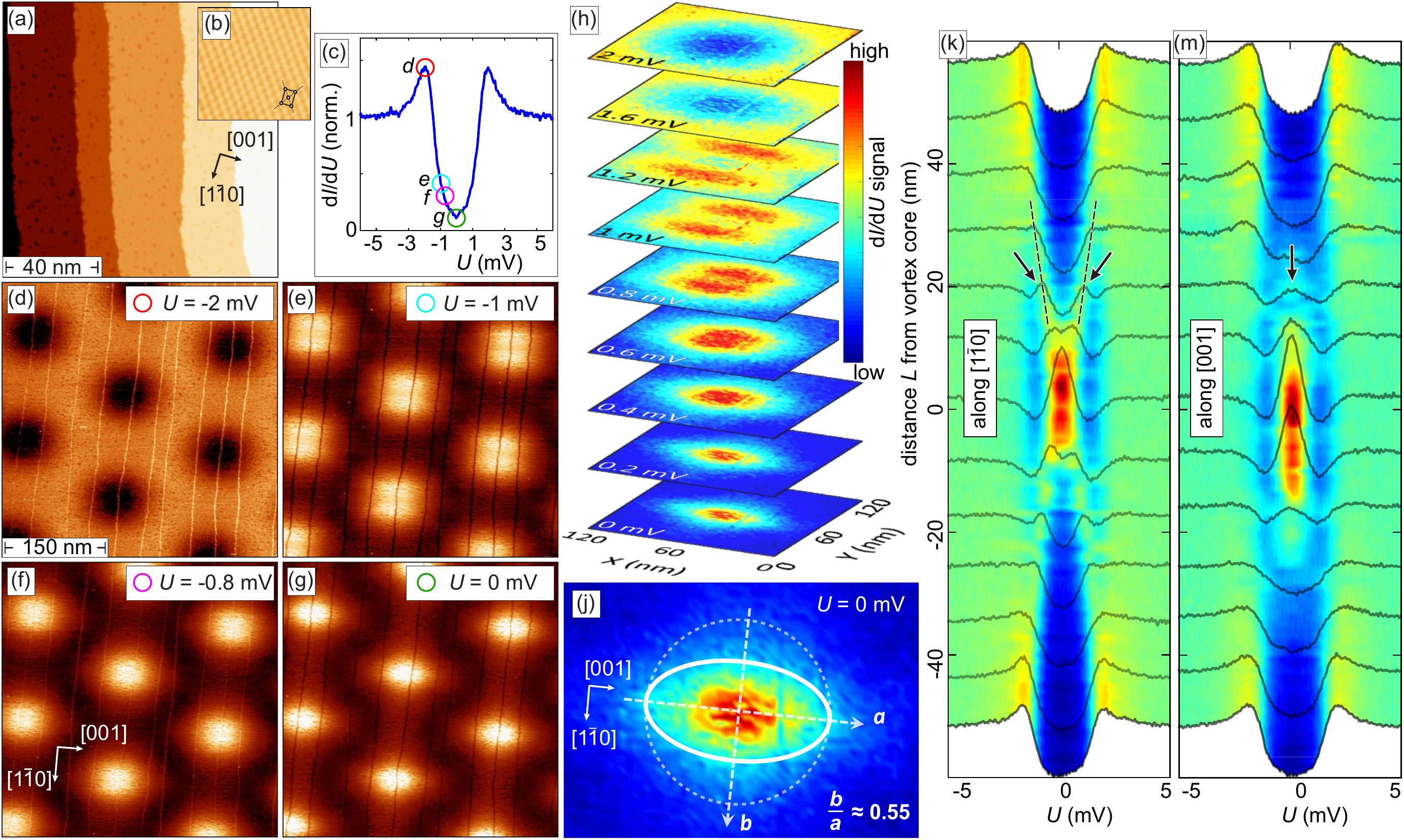}
	\caption{(a) Topography of clean Nb(110).
	(b) Atomic resolution scan with indicated unit cell.
	(c) Tunneling spectrum of the superconducting gap.
	(d)-(g) Constant-separation $\mathrm{d}I/\mathrm{d}U$ maps 
	taken at the voltages marked by colored circles in (c) at $\mu_0H = 100$\,mT.  	
	Whereas the vortices appear round-shaped at a bias voltage close to the edge of the gap,
	they deform into ellipses when approaching zero bias.
	(h) Stacked representation showing the bias-dependent LDOS inside the vortex in the energy window from 0 to 2\,mV.
	(j) Fermi level ($U = 0$\,V) $\mathrm{d}I/\mathrm{d}U$ map
	revealing an elliptical shape elongated along the $[001]$ direction with an eccentricity $b/a = 0.55 \pm 0.1$.
	(k),(m) Waterfall plot of tunneling spectra taken along the lines labeled $b$ and $a$ in (j), respectively.
	With increasing distance from the vortex core along the $[1\overline{1}0]$ direction the zero-bias peak splits into two ridges, 
	resulting in the appearance of an X-shaped feature as indicated by two arrows and dashed lines in (k).
	In contrast, along the [001] direction only a vanishing of the peak without any splitting is observed (m).
	Scan parameters: (a)-(m) $I_{\rm set} = 200$\,pA; $T = 1.4$\,K (a) $U_{\rm set} = -1$\,V; (c) $U_{\rm set} = -10$\,mV;
	(d)-(g)  $U_{\rm set} = -100$\,mV; (h)-(j)  $U_{\rm set} = -10$\,mV; (k)-(m)  $U_{\rm set} = -7$\,mV.}
	\label{fig:Nb}	
\end{figure*}
{\em Experimental procedures---}
STM and STS measurements are performed in a home-built setup (base temperature $T = 1.4$\,K) which is equipped
with a superconducting magnet capable of delivering a magnetic field up to $\mu_0H = 12$\,T along the surface normal.
Spectroscopic measurements are conducted by a lock-in technique
with $U_{\textrm{mod}} = 0.1\div0.2$\,meV at a modulation frequency $f = 890$\,Hz. 
Nb(110) was cleaned by a series of high temperature flashes up to $T_{\rm fl} \approx 2400^{\circ}$C.
As described in detail in Ref.\,\onlinecite{Odobesko2019}, lower flash temperatures
result in two intermediate and less-ordered oxygen reconstructions, i.e.,
NbO$_x$ phase-I at $T_{\rm fl} \le1800^{\circ}$C and NbO$_x$ phase-II
for $2000^{\circ}$C\,$\le T_{\rm fl} \le 2300^{\circ}$C \cite{SupplMat}.

{\em Experimental results---}
Figure \ref{fig:Nb}(a) shows the constant-current STM topography image of a clean Nb(110) surface
with occasional oxygen-reconstructed patches which appear as dark spots ($< 10$\% of the surface area).
The atomic resolution scan reported in Fig.\,\ref{fig:Nb}(b) confirms the expected lattice constant $a_{[001]} = 330$\,pm.
Nb is a type-II superconductor with $T_{\rm c}^{\rm Nb} = 9.2$\,K which---if exposed
to an external magnetic field---forms an Abrikosov lattice of normal conducting vortices.
Tunneling spectra taken far from any vortex show a well-defined superconducting gap with $\Delta = 1.53$ meV [Fig.\,\ref{fig:Nb}(c)].
We carefully investigate the spatial variation of the electronic structure around vortex cores
by mapping the differential conductance $\mathrm{d}I/\mathrm{d}U$ at various energies $E = eU$ within the superconducting gap.
To avoid set-point--related artifacts, all measurements are performed by stabilizing the STM tip
at a set-point bias voltage corresponding to an energy far within the normal metallic regime, i.e., at $U_{\rm set} = -100$\,mV.
After recording the \textit{z}-trace along one scan line at this bias voltage
the feedback is switched off, the bias voltage is set to measurement parameters,
and the tip is approached towards the surface by a fixed value $\Delta z = 180$\,pm.
Afterwards the $\textrm{d}I/\textrm{d}U$ signal is measured along the (shifted) \textit{z}-trace.
This procedure increases the tunneling current and thereby improves the signal-to-noise ratio.
At the same time, it guarantees that the $\mathrm{d}I/\mathrm{d}U$ maps
presented in Figs.\,\ref{fig:Nb}(d)-(g) are measured at a constant tip--sample separation.

Figures\,\ref{fig:Nb}(d)-(g) show the Abrikosov lattice formed on clean Nb(110) in a magnetic field $\mu_0H = 100$\,mT
at four selected tunneling voltages indicated by colored circles in Fig.\,\ref{fig:Nb}(c).
Some qualitative differences regarding the appearance of the vortices are immediately evident.
At $U = -2$\,mV [Fig.\,\ref{fig:Nb}(d)], i.e., at an energy which corresponds to the maximum DOS at the coherence peak,
the vortices appear as dark circular regions with a diameter of about 50\,nm.
As we approach the Fermi level, the vortices split along the $[1\overline{1}0]$ direction [Fig.\,\ref{fig:Nb}(e)],
and eventually deform into ellipses [Fig.\,\ref{fig:Nb}(f)-(g)].
Furthermore, due to the vanishing of the superconducting gap and the enhanced quasiparticle LDOS inside the vortex, 
the vortices appear brighter than the surrounding superconducting area in Fig.\,\ref{fig:Nb}(e)-(g).

A more detailed, stacked representation of the energy-dependent vortex shape
is presented in Fig.\,\ref{fig:Nb}(h) for positive bias voltages.
As can be expected for particle--hole symmetric excitations in superconductors,
the behavior is equivalent to the scenario described in panels (d)--(g) for negative bias.
Again, the vortex appears as a circular region of reduced LDOS at the coherence peaks ($U = 2$\,mV).
As the voltage is reduced the LDOS intensity becomes (i) larger inside the vortex than in the surrounding superconducting region,
it (ii) splits along $[1\overline{1}0]$ direction (1.2\,mV $\ge U \ge 0.7$\,mV),
and eventually (iii) elongates along the $[001]$ direction ($U \le 0.7$\,mV).
Fig.\,\ref{fig:Nb}(j) presents the $\mathrm{d}I/\mathrm{d}U$ map of a vortex measured at the Fermi level ($U = 0$\,mV).
Its shape clearly deviates from a circle (hatched line) and rather corresponds to an ellipse (thick white line)
with the semi-major $a$ and semi-minor $b$ axes oriented along the $[001]$ and $[1\overline{1}0]$ direction, respectively.
The ratio $b/a$ amounts to $\approx 0.55$.
We would like to emphasize that the orientation of the ellipse does not depend on the presence of defects,
such as step edges, and stays the same at higher magnetic fields.

\begin{figure}[t]
	\includegraphics[width=1\columnwidth]{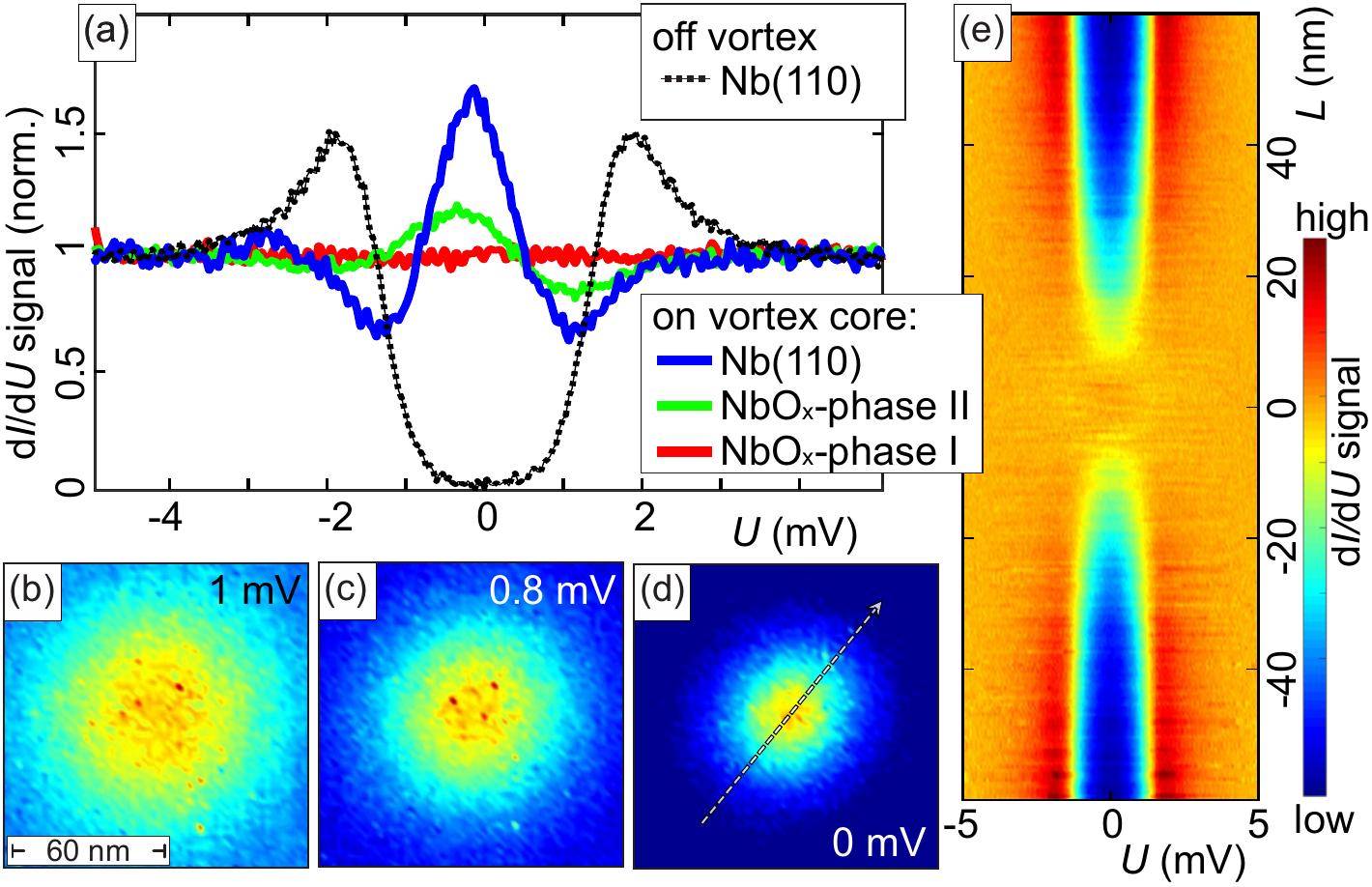}
	\caption{(a) Tunneling spectra measured on different Nb surfaces in an external magnetic field:
	(black) the fully gapped tunneling spectrum of clean Nb(110) far away from any vortex;
	(blue) intense zero-bias peak in the center of the vortex on clean Nb(110);
	(green) less intensive peak in a vortex core on NbO$_x$ phase-II;
	(red) Ohmic behavior in vortex core on NbO$_x$ phase-I. 	
	Stabilization parameters: $U_{\rm set} = -7$\,mV, $I_{\rm set} = 100$ pA.
	Panels (b)-(d) show differential conductance $\mathrm{d}I/\mathrm{d}U$ maps of a single vortex on NbO$_x$ phase-I
	imaged at (b) $U = 1$\,mV, (c) $U = 0.8$\,mV, and (d) $U = 0$\,mV.
	(e) Tunneling spectra measured along the line in (d). No zero-bias feature is observed.
	Stabilization parameters: $U_{\rm set} = -5$\,mV, $I_{\rm set} = 100$\,pA, $\mu_0H = 100$\,mT, $T = 1.5$\,K. 	}
	\label{fig:Spectra}
\end{figure}
Figures\,\ref{fig:Nb}(k) and (m) show the spatial evolution of tunneling spectra
measured along the axes $b$ and $a$. 
The spectrum measured at the vortex center is characterized by a strong zero-bias peak,
representing CdGM states which are thermally blurred into a single peak \cite{Hess1990,Hess1991}. 
However, along the two high-symmetry axes we observe qualitatively different transitions
from the peaked spectrum measured at the vortex center to the fully gapped spectrum far away.
Whereas the  zero-bias peak splits in energy into two ridges if measured along the $[1\overline{1}0]$ direction
[marked by two black arrows and hatched lines in Fig.\,\ref{fig:Nb}(k)], 
a behavior expected for CdGM state \cite{Gygi1991,Hayashi1997},
no such splitting is detected along $a$, i.e., the major axes along the $[001]$ direction [Fig.\,\ref{fig:Nb}(m)].

A detailed investigation of the electronic structure inside vortices observed on 
ordered clean Nb(110), the less-ordered NbO$_x$ phase-II, or the disordered NbO$_x$ phase-I \cite{Odobesko2019} 
reveals that the existence and shape of ZBP features critically depends on the surface quality.
Fig.\,\ref{fig:Spectra}(a) presents tunneling spectra measured in vortex cores on all three surfaces.
For comparison, the black curve in Fig.\,\ref{fig:Spectra}(a) shows the spectrum 
of superconducting clean Nb(110) far away from a vortex, clearly presenting a superconducting gap.
Evidently, the spectral features observed around zero bias exhibit marked, surface structure-dependent differences.
Whereas the ZBP intensity measured on clean Nb(110) (blue) clearly exceeds the normal conducting DOS, 
indicating the presence of CdGM states in the vortex core, the ZBP is less pronounced on the NbO$_x$ phase-II (green) 
and essentially absent for the NbO$_x$ phase-I (red). 

This reduction of the ZBP is accompanied by a more isotropic appearance of vortices, 
as evidenced by the $\textrm{d}I/\textrm{d}U$ map of a vortex 
on the NbO$_x$ phase-I presented in Fig.\,\ref{fig:Spectra}(b-d),  
which remains circular throughout the entire energy range within the superconducting gap.  
Furthermore, the tunneling spectra presented in Fig.\,\ref{fig:Spectra}(e) which were taken along the line in (d) 
across the vortex reveal the absence of any feature potentially related to bound states. 
These findings suggest that the origin of the anisotropy observed for vortices on the clean Nb(110) surface 
results from the presence of the CdGM states. 
Their absence on the less ordered NbO$_x$ phase-I is consistent with 
a cross-over from the clean to the dirty limit \cite{Renner1991}. 
If we assume for the clean Nb(110) surface a mean free path close to the bulk value, 
$\ell^{\rm Nb} \geq 100$\,nm \cite{NeilW.Ashcroft1976}, 
it clearly exceeds the coherence length $\xi^{\rm Nb} \approx 38$\,nm. 
Since the oxygen-reconstructed surfaces of Nb are less ordered, 
additional scattering of quasiparticles may substantially decrease the mean free path and quench the CdGM states.

{\em Discussion---}
Anisotropic vortices were first observed by Hess \textit{et al}.\ on superconducting NbSe$_2$ \cite{Hess1990, Hess1991}.
Three scenarios were discussed as potential reasons: (i) vortex--vortex interactions, (ii) anisotropic pairing, 
and (iii) an anisotropy of the Fermi surface \cite{Hayashi1996, Hayashi1997, Gygi1991}.
The effect of vortex--vortex interactions becomes relevant for spacings
comparable with or smaller than the London penetration depth $\lambda$.
Under this condition the stray field-mediated interaction between a given vortex
and adjacent flux lines may cause a significant distortion of its shape \cite{Gygi1991}.
Furthermore, as the spacing approaches the characteristic coherence length $\xi$,
the hybridization of the nearest neighbor quasiparticle wave functions may also affect the LDOS \cite{Melnikov2009}.
However, both $\lambda^{\rm Nb} \approx 39$\,nm and $\xi^{\rm Nb} \approx 38$\,nm 
are at least four times smaller than the Abrikosov lattice constant ($\approx 160$\,nm) observed at $\mu_0 H = 100$\,mT.
Therefore, we rule out that scenario (i) is responsible for the observed elliptical shape in Nb.

The two-fold rotational symmetry of (110) surfaces of body-centered cubic crystals, such as Nb,
certainly implies  inequivalent electronic properties along the two high-symmetry directions [001] and [$1 {\overline 1} 0$].
Since both scenarios, (ii) and (iii), are intimately related to the surface electronic structure of the respective superconductor, 
either of the two effects may---in principle---be responsible for the anisotropic dispersion of bound states inside the vortex. 
However, experiments performed on crystalline Nb planar tunnel junctions suggest that the anisotropy of the superconducting gap
does not exceed 10\% of the average value ($\Delta = 1.53$ meV) \cite{MacVicar1968, MacVicar1970},
insufficient to explain the strong anisotropy observed in Fig.\,\ref{fig:Nb}(j).

\begin{figure}[t]
	\includegraphics[width=1\columnwidth]{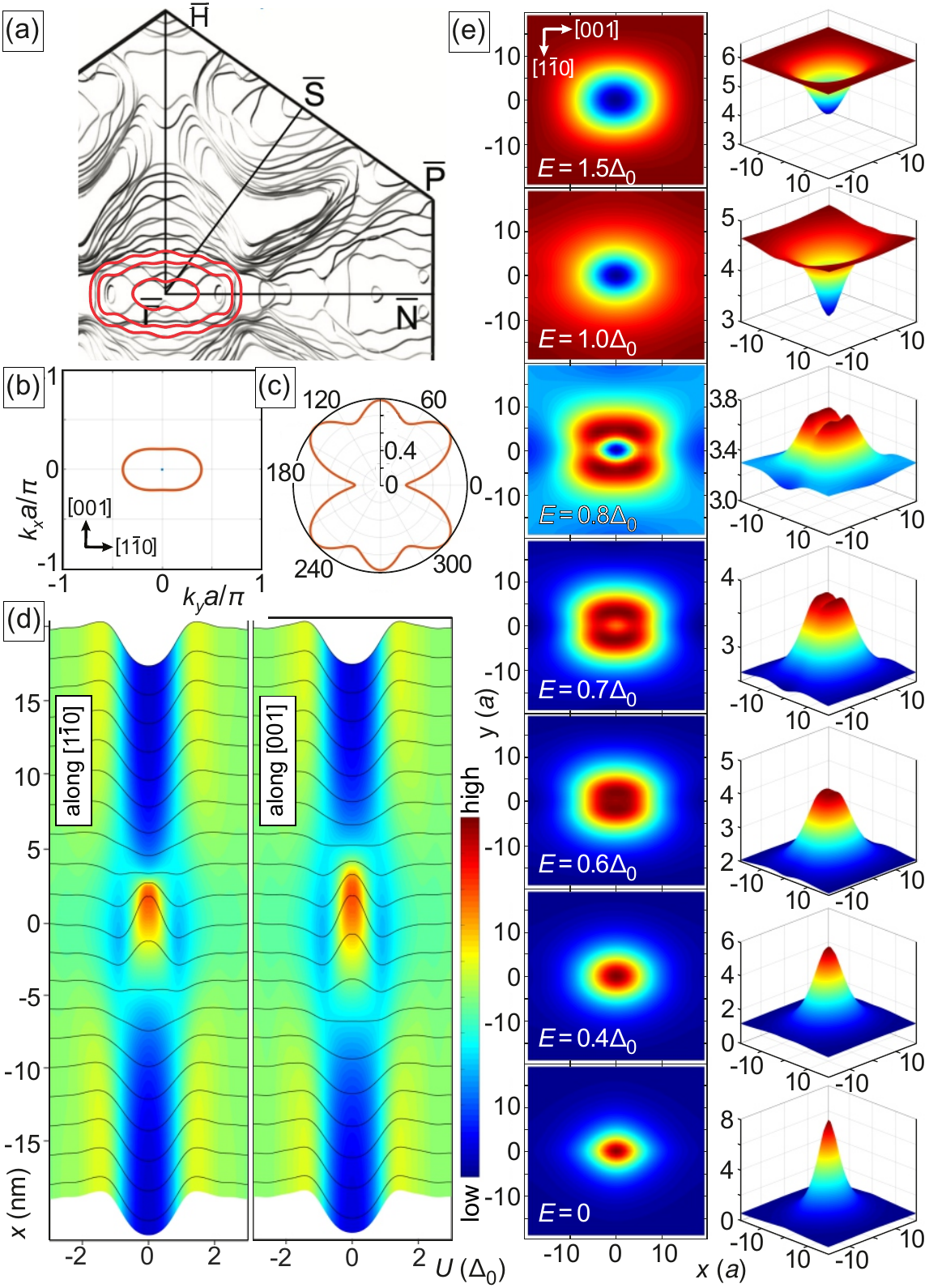}
	\caption{(a) Calculated Fermi surface of Nb(110).  
		Red lines denote pockets around the $\overline{\Gamma}$ point. 
		(b) Hypothetical Fermi surface and  (c) Fermi velocity of the dominant Fermi contour. 
		(d) Calculated LDOS along two principal directions crossing the vortex core. 
		(e) Excitation energy-dependent LDOS profiles after thermal broadening with $k_BT \simeq 0.26\Delta_0$}.
	\label{fig:Theory}
\end{figure}

To explain the experimental observations, we calculate the Fermi surface of clean Nb(110) 
using the plane-wave-based VASP~\cite{vasp1,vasp2} code 
within the projector augmented-wave method~\cite{blo,blo1}. 
The generalized gradient approximation of Perdew-Burke-Ernzerhof~\cite{PBE,PBEerr} is used for the exchange correlation. 
Details of theory methods and examples of the good agreement between theoretical 
and measured LDOS can be found in Ref.~\onlinecite{Odobesko2019}. 
The calculated Fermi surface of the Nb(110) surface is plotted in Fig.\,\ref{fig:Theory}(a).
Multiple Fermi surface pockets centered around the various high-symmetry points 
$\overline{\Gamma}$, $\overline{\mathrm{N}}$, $\overline{\mathrm{S}}$, 
and $\overline{\mathrm{H}}$ can be recognized \cite{Odobesko2019}.
Since STM is most sensitive to electronic states in vicinity to the $\overline{\Gamma}$ point of the SBZ, 
with other states decaying exponentially with increasing $\bf{k}_{\parallel}$ \cite{Tersoff1985, Heinze1998}, 
we expect that the stadium-shaped contours (red) in Fig.\,\ref{fig:Theory}(a) contribute most to the measured signal.
Therefore, we employ similar contours to approximate the Fermi surface within an effective model described in the following.

In this model, we consider an electron gas on a 2D square lattice, with different effective masses, 
$m_x/m_y\approx 0.47$, in two principal directions and with an $s$-wave pairing interaction. 
In order to mimic a truly anisotropic band structure, we further consider an isotropic spin-orbit interaction, 
which can be either of Rashba or Dresselhaus type\ \cite{Dresselhaus55PR,Rashba60SPSS}. 
This spin-orbit interaction results in two split Fermi surfaces. 
The larger Fermi surface takes a stadium shape and carries a strong anisotropic Fermi velocity, 
see Fig.\,\ref{fig:Theory}(b), while the smaller one is circular. 
We set the chemical potential at a value where the stadium-shaped Fermi surface dominates. 
Additionally, we apply a quantized magnetic flux and impose magnetic periodic boundary conditions on the lattice. 
Then, we solve the resulting BdG equations self-consistently on the lattice and calculate the LDOS 
around a vortex core at a low temperature set to $k_{\rm B}T\simeq0.26\Delta_0$, 
where $\Delta_0$ is the pairing potential in the absence of the magnetic field. 
We present the results in Fig.\,\ref{fig:Theory}(d-e) 
and provide more details about the simulation in Ref.\ \onlinecite{SupplMat}. 
At low excitation energies, the LDOS is enhanced at the vortex core with an elliptical shape. 
This changes to a coffee-bean shape at intermediate excitation energies 
and to a round shape at high excitation energies, 
in excellent qualitatively agreement with experimental observations.

Our analysis confirms that the anisotropy of the Fermi surface is carried over 
to the spatial structure of the excited states within the vortices. 
Depending on details of the anisotropy of the Fermi surface and the resulting Fermi velocities 
and the spin-orbit coupling, a distinct anisotropy of the CdGM states can be observed 
even in a system with an isotropic order parameter.

{\em Conclusion---}
In conclusion, we have presented STS studies of vortex bound states 
of the clean Nb(110) surface in an external magnetic field.
Our results reveal that the vortex shape in $\mathrm{d}I/\mathrm{d}U$ maps depends on the energy,  
transforming from circular to a coffee-bean shape, and finally to an ellipses when approaching $E_{\rm F}$. 
We provided evidence that the observed energy-dependent spatial distribution 
originates from Caroli-de Gennes-Matricon states which inherit their anisotropy from the Nb(110) Fermi surface. 
By oxidizing the Nb(110) surface we could trigger the transition from the clean to the dirty limit, 
accompanied by the complete disappearance of any anisotropy in the LDOS 
due to the quenching of the vortex bound states.
These findings will be useful in the future for distinguishing trivial from topological vortex bound states by, 
for example, their spatial distribution or by introducing disorder.

\begin{acknowledgments}
This research was supported by the DFG (through SPP1666 and SFB 1170 ``ToCoTronics'', 
the W\"urzburg-Dresden Cluster of Excellence ct.qmat, EXC2147, project-id 390858490, 
and the Elitenetzwerk Bayern Graduate School on ``Topological Insulators".
S.J.H.\ and S.H.\ thank the Norddeutscher Verbund f{\"u}r Hoch- und H{\"o}chstleistungsrechnen (HLRN) for providing computational resources.
\end{acknowledgments}

\bibliography{Vortex_bib_12}

\end{document}